\documentclass[a4paper,11pt]{article}
\usepackage{latexsym}
\usepackage{rotating}
\usepackage{amsmath}
\author{Wen-Xiu~Ma\footnote{Email: mawx@cityu.edu.hk} \\
Department of Mathematics, City University of Hong Kong,\\
Kowloon, Hong Kong, China}
\title {A Bi-Hamiltonian Formulation for Triangular
Systems \\ by Perturbations}
\date{\nonumber}
\begin{document}

\setlength{\baselineskip}{18pt}
\maketitle

\
\vskip 5mm
\begin{abstract}
  
A bi-Hamiltonian formulation is proposed 
for triangular systems resulted by perturbations
around solutions, from which infinitely many symmetries and 
conserved functionals of triangular systems can be explicitly constructed,
provided that one operator of the Hamiltonian pair is invertible. 
Through our formulation, four examples of triangular systems 
are exhibited, which also show that bi-Hamiltonian systems 
in both lower dimensions and higher dimensions
are many and varied.
Two of four examples give local $2+1$ dimensional bi-Hamiltonian systems and 
illustrate that multi-scale perturbations 
can lead to higher-dimensional bi-Hamiltonian systems.

\end{abstract}

\newcommand{\R}{\mbox{\rm I \hspace{-0.9em} R}}

\def \ba {\begin{array}}
\def \ea {\end{array}}
\def\bea{\begin{eqnarray}}
\def\eea{\end{eqnarray}}

\newcommand{\eqnsection}{
   \renewcommand{\theequation}{\thesection.\arabic{equation}}
   \makeatletter
   \csname $addtoreset\endcsname
   \makeatother}
\eqnsection

\newtheorem{lemma}{Lemma}[section]
\newtheorem{theorem}{Theorem}[section]
\newtheorem{definition}{Definition}[section]

\section{Introduction}
\setcounter{equation}{0}

The bi-Hamiltonian formulation is a great success in the Hamiltonian theory
of differential equations \cite{Magri-JMP1978}.
It has attracted the attention of a wide audience within both the
mathematical community and the physical community due to its importance 
in producing symmetries and conserved functionals,
and has already become one of active research directions in the field
of soliton theory and integrable systems. 

In this paper, we are concerned with the bi-Hamiltonian formulation
of triangular systems resulted by various perturbations around 
solutions, specific systems of which were furnished
in \cite{CaseR-JMP1981,LakshmananT-JMP1985,MaF-CSF1996}. Such triangular systems 
provide candidates of integrable couplings for given 
integrable systems 
\cite{Fuchssteiner-book1993,Ma-MAA1998}. A general triangular system reads as
\begin{equation} \left\{\begin{array} {l} 
u_t=K(u)=K(u,\cdots, u^{(k)}),
\vspace{2mm} \\
v_t=S(u,v)=S(u,v,\cdots, u^{(l)},v^{(l)}),
\end{array} \right. \end{equation}
where $u=u(t,x),\ v=v(t,x)$, and $u^{(n)}$ and $ v^{(n)}$ are 
derivatives with respect to the spatial variable $x$. 
Such a concrete example by a first-order perturbation is given by 
\begin{equation} \left\{ \ba {l}u _{t}=K(u), \vspace{2mm}\\
v _{t}=K'(u)[v], \ea \right.
\label{symmetrysystem} \end{equation}
where $K'(u)[v]$ denotes the Gateaux derivative of $K(u)$ at a direction 
$v$, i.e.,
\[K'(u)[v]=\frac {\partial }{\partial \varepsilon } \Bigl.\Bigr|_{\varepsilon =0}
K(u+\varepsilon v).
 \]
A mathematical structure called the perturbation bundle 
has been established in \cite{Fuchssteiner-book1993} to study its integrable properties.
Note that the second component of the above system is just the linearized system 
of the original system $u_t=K(u)$, and thus the symmetry 
problem leads to a triangular system together with the original system,
which also shows the importance of studying triangular systems
(see \cite{Ma-MAA1998} for more discussion). 
Other similar examples of specific triangular systems were presented by means of 
perturbations in \cite{CaseR-JMP1981,LakshmananT-JMP1985,MaF-CSF1996}.
However, there is no discussion on the bi-Hamiltonian formulation
of general triangular systems by perturbations, even the specific systems mentioned above.       

With a view to exposing integrability, 
we would like to answer whether there exists  
any bi-Hamiltonian formulation  
%what they are if there exist
for triangular systems resulted by various perturbations around 
solutions. It will be shown that 
%if we start from a bi-Hamiltonian system, 
a bi-Hamiltonian formulation of the resulting triangular systems 
can be inherited from an original bi-Hamiltonian system.
The general formulation allows us to present 
various examples of bi-Hamiltonian systems, in both $1+1$ and $2+1$ dimensions.

The paper is organized 
as follows. In Section 2, we shall choose perturbed 
systems of given bi-Hamiltonian systems as starting systems
and introduce our triangular systems 
by using perturbations around solutions of starting 
systems, which contain many 
special triangular systems in 
\cite{CaseR-JMP1981,LakshmananT-JMP1985,MaF-CSF1996}. 
Then in Section 3, 
a bi-Hamiltonian formulation for the resulting triangular systems 
will be proposed, based on the bi-Hamiltonian formulation of  
starting systems.
In Section 4, we will go on to exhibit
four examples of triangular systems 
through the general bi-Hamiltonian formulation, which also show that bi-Hamiltonian systems in both $1+1$ and $2+1$ dimensions
are many and varied.
Two of four examples give local $2+1$ dimensional bi-Hamiltonian systems and 
illustrate that multi-scale perturbations 
can lead to higher-dimensional bi-Hamiltonian systems.
Finally in Section 5,
some concluding remarks will be given.

\section{Triangular systems by perturbations}
\setcounter{equation}{0}

{\bf 2.1 Bi-Hamiltonian systems:}

Assume that we have a bi-Hamiltonian system
\begin{equation}
u_t=K(u)=J\frac {\delta \tilde H_1}{\delta u}=
M\frac {\delta \tilde H_0}{\delta u}
,\ \tilde H_0=\int H_0\,dx,\ \tilde H_1=\int H_1\,dx,
\label{originalsolitonsystem}
\end{equation}
where $J$ and $M$ constitute a Hamiltonian pair
(see \cite{GelfandD-FAA1979,FokasF-LNC1980,Blaszak-book1998}
for more information), $t$ is a single variable but $x$ can be
a single or vector variable.
If one operator of the Hamiltonian pair is 
invertible, we can have infinitely many symmetries 
$\{K_n \}_{n =0}^\infty$ and conserved functionals 
$\{ {\tilde H}_n \}_{n =0}^\infty$, which can be explicitly computed through
\begin{equation}
\left\{\ba {l} 
K_n =\Phi ^{n -1}K(u)=(MJ^{-1}) ^{n -1}K(u),\ n\ge 1,\vspace{2mm}\\
{\tilde H}=\int H_ndx,\  H_{n }= \int_0^1\langle u,(J^{-1}\Phi ^{n -1}K)(\lambda u)\rangle
d\lambda ,\ n \ge 0, \ea \right.
\label{generaleschemeforK_nH_n}
\end{equation}
where $\langle \cdot,\cdot\rangle $ denotes the standard inner product of the corresponding 
Euclidean space.
Moreover, they are related through the bi-Hamiltonian formulation 
\cite{Olver-book1993}
\begin{equation}
K_n =J\frac {\delta \tilde H_{n }}{\delta u}
=M\frac {\delta \tilde H_{n -1}}{\delta u},\ 
%\tilde H_n =\int H_n \, dx,\ 
 n \ge 1.
\end{equation}
Fuchssteiner and Fokas \cite{FuchssteinerF-PD1981} discovered 
an important fact that 
when $J$ and $M$ constitute a Hamiltonian pair and $J$ is invertible,
the operator $\Phi =MJ^{-1}$ is hereditary \cite{Fuchssteiner-NATMA1979},
i.e., 
\[ \Phi'[\Phi X] Y - \Phi \Phi'[X]Y =\Phi'[\Phi Y] X - \Phi \Phi'[Y]X   \]
holds 
for any vector fields $X$ and $Y$, where $\Phi'[X]$ denotes the Gateaux
derivative
\[\Phi'[X]= \frac {\partial }{\partial \varepsilon } \Bigl.\Bigr|_{\varepsilon =0}
\Phi (u+\varepsilon X).
\]
This condition is actually equivalent to an invariance of Lie derivative of $\Phi$ 
(see, for example, \cite{Ma-JPA1990,Ma-SCA1991}).
%provided that $J$ and $M$ constitute a Hamiltonian pair. 
It is the hereditaryness of $\Phi$ that gives rise to an explanation
why soliton systems come in hierarchies.

{\bf 2.2 Perturbed systems:}

Let us now 
choose a perturbed system with a perturbation parameter $\varepsilon $:
\begin{equation} u_t=K^{\textrm{{\small per}}}(u):=
\sum_{j=0}^{m} \alpha _j \varepsilon ^j K_{i_j}(u)=
J\frac {\delta \tilde H^{\textrm{{\small per}}}_1} {\delta u} =M
\frac {\delta \tilde H^{\textrm{{\small per}}}_0} {\delta u},
\label{startingPS}
\end{equation}
where $m\ge 0$, the $\alpha _j$ are arbitrary constants, the $i_j$
are arbitrary natural numbers (which means to take arbitrary vector fields $
 K_{i_j}(u)$ from $\{K_n \}_{n =1}^{\infty}$), and two Hamiltonian functionals read
\begin{equation} \tilde H^{\textrm{{\small per}}}_0=
\int H^{\textrm{{\small per}}}_0\, dx,\ 
 H^{\textrm{{\small per}}}_0=
\sum_{j=0}^m\alpha _j \varepsilon ^j  H_{i_j},\ 
 \tilde H^{\textrm{{\small per}}}_1=
\int H^{\textrm{{\small per}}}_1\, dx,\ 
 H^{\textrm{{\small per}}}_1=
\sum_{j=0}^m\alpha _j \varepsilon ^j  H_{i_j+1}.\label{originalH_01per}
\end{equation}
This system (\ref{startingPS}) is called a starting system, which is nothing but  
a generalized system of the original bi-Hamiltonian system (\ref{originalsolitonsystem}).
Following the general scheme shown in (\ref{generaleschemeforK_nH_n}), 
we have infinitely many symmetries and conserved functionals
\begin{equation} \left \{ \begin{array} {l} 
{\displaystyle K^{\textrm{{\small per}}}_n:=\Phi ^{n-1}K^{\textrm{{\small per}}}=
\sum_{j=0}^{m} \alpha _j \varepsilon ^j K_{i_j+n-1}(u),\ n\ge 1,}
\vspace{2mm} \\
{\tilde H}^{\textrm{{\small per}}}_n:={\displaystyle \int {H}^{\textrm{{\small per}}}_ndx,
\ {H}^{\textrm{{\small per}}}_n= 
\sum_{j=0}^{m} \alpha _j \varepsilon ^j H_{i_j+n}(u),\ n\ge 0,}\end{array}
\right.
\label{K_n^{per}H_n^{per}}
\end{equation}
for the starting system (\ref{startingPS}), since we can directly check that 
\[ K^{\textrm{{\small per}}}_n
=J\frac {\delta {\tilde H}_n ^{\textrm{{\small per}}}} {\delta u}=
M  \frac {\delta {\tilde H}_{n-1} ^{\textrm{{\small per}}}}{\delta u},\ n\ge 1.
 \]

{\bf 2.3 Triangular systems:}

For any integers $N\ge 0$ and $r\ge 0$, take a perturbation series:
\begin{equation}
\hat{  u}_N=\sum_{i=0}^N \varepsilon  ^i\eta _i , \
\eta _i=\eta _i(y_0, y_1, y_2,\cdots,y_r,t),
\label{multiplescaleperturbationseries}
\end{equation}
where $y_i=\varepsilon ^i x,\ 0\le i\le N,$ are all slow variables. 
Now we make a perturbation around solutions of the starting system 
(\ref{startingPS})
and observe the $N$-th order perturbation system 
\[\hat {u}_{Nt}=K^{\textrm{{\small per}}}(\hat {u}_N)+\textrm{o}(
\varepsilon ^N) ,\]
where  
$u$, $\eta _i$, $0\le i\le N$, are supposed to be column vectors of the same dimension. 
By the Taylor expansion, this leads to an equivalent and bigger system 
\begin{equation}
\eta _{it}=
\frac{1}{i!}\frac {\partial ^i}{\partial \varepsilon ^i}
K ^{\textrm{\small {{per}}}}
(\hat u_N)\Bigl.\Bigr |
_{\varepsilon =0} = 
\displaystyle{\sum_{j=0}^{\textrm{\small {{min}}}
(m,i)}\frac {\alpha _j}
{(i-j)!}\frac {\partial ^{i-j}}{\partial \varepsilon ^{i-j}}}
K_{i_j}(\hat {u}_N)\Bigl.\Bigr |
_{\varepsilon =0},\ 0\le i\le N. \label{triangularSysByPer} \end{equation}
For brevity, we rewrite it as a concise form 
\begin{equation} 
\hat {\eta } _{Nt}=(\textrm{per}_NK^{\textrm{{\small per}}})(\hat {\eta }
_N)
%=\hat {K}^{\textrm{{\small per}}}_N(\hat {\eta }_N)
= 
((\textrm{per}_N {K}^{\textrm{{\small per}}})_0^T,\cdots ,(\textrm{per}_N {K}^{\textrm{{\small per}}})_N^T)^T ,
\  \hat{\eta }_N=
(\eta _0^T,\cdots ,\eta_N^T)^T,
\end{equation}
where $T$ denotes the matrix transpose. 
Noting that 
\[ \hat u_N =\hat u_i 
 + \varepsilon ^{i+1} \sum_{j=0}^{N-i-1}
\varepsilon ^{j}\eta _{j+i+1}  ,\ \hat u_i=\sum _{j=0}^i \varepsilon ^j \eta _j
, \ 0\le i\le N-1,
\]
an application of the Taylor expansion tells us that 
\[  (\textrm{per}_N K^{\textrm{{\small per}}})_i = \frac 1{i!}\frac {\partial ^i}
{\partial \varepsilon ^i} K^{\textrm{{\small per}}}(\hat u_N)
\Bigl.\Bigr|_{\varepsilon =0}  = 
\frac 1{i!}\frac {\partial ^i}
{\partial \varepsilon ^i} K^{\textrm{{\small per}}}(\hat u_i)
\Bigl.\Bigr|_{\varepsilon =0}
,\ 0\le i\le N-1,
 \]
and thus the perturbation system (\ref{triangularSysByPer})
is triangular, i.e., the $(i+1)$-th component $\eta _{it}=
(\textrm{per}_N K^{\textrm{{\small per}}})_i $ just involves 
the first $i+1$ dependent variables $\eta _0,\cdots ,\eta _i$
but no the other dependent variables $\eta _{i+1},\cdots ,\eta _N$.

In our formulation,
the superscript ``per" denotes the perturbed objects such as 
the perturbed tensor fields and the perturbed functionals
as in (\ref{startingPS}) and (\ref{originalH_01per}),
but the prefix ``per$_N$" means the perturbation resulting from 
the $N$-th order perturbation 
(\ref{multiplescaleperturbationseries}) of the dependent variable $u$.
%We also emphasize that 
The small parameter $\varepsilon $ 
is involved in both the starting system (\ref{startingPS}) and 
the perturbation series (\ref{multiplescaleperturbationseries}), but  
there is no relation among three integers $m$, $N$, and $r$ that we need to take 
in the starting system (\ref{startingPS}) and 
the perturbation series (\ref{multiplescaleperturbationseries}).
%These will offer various choices of our triangular systems. 
This demonstrates diversity to formulate our triangular systems.
If we take a special choice of $\alpha _0=1$ and $K_{i_0}=K$ in our construction, 
the triangular system
(\ref{triangularSysByPer}) becomes a coupling system of $u_t=K(u)$, 
because its first component is $\eta _{0t}=K(\eta _0).$ This paves a way for constructing
integrable couplings of given integrable systems \cite{MaF-CSF1996,Ma-MAA1998}.
If a starting system is particularly chosen as 
\begin{equation}
u_t=K^{\textrm{{\small per}}}(u)=K(u)+\alpha \varepsilon   K(u), \ \alpha =\textrm{const.},
\end{equation}
the following specific triangular system 
\begin{equation}
\left \{ \ba {l} \eta _{0t}=K(\eta _0), \vspace{2mm}\\
\eta _{1t}=K'(\eta _0)[\eta _1]+  \alpha K(\eta _0),
\ea  \right. \label{specificIC}
\end{equation}
will be engendered upon making a first-order perturbation. 
This system looks simple, but it generalizes the triangular 
system (\ref{symmetrysystem})
resulting from the symmetry problem.
The main objective of this paper is to propose a bi-Hamiltonian formulation for 
the triangular systems determined by (\ref{triangularSysByPer}), 
%which will lead to new bi-Hamiltonian systems in both $1+1$ and $2+1$ dimensions
which contain two
specific interesting triangular systems (\ref{symmetrysystem})  
and (\ref{specificIC}).

\section{Bi-Hamiltonian formulation}
\setcounter{equation}{0}

For now on, we focus on the establishment of a bi-Hamiltonian formulation
for the triangular systems determined by (\ref{triangularSysByPer}).
We would actually like to show that 
a bi-Hamiltonian formulation of the resulting triangular systems 
can be inherited from an original bi-Hamiltonian system (\ref{originalsolitonsystem}).

To the end, let us first introduce a new Hamiltonian pair:
\begin{equation}
(\textrm{per}_N J)(\hat{\eta  }_N)\equiv \hat{J}_N (\hat {\eta}_N  )
\ \ \textrm{and}\ \ 
 (\textrm{per}_N M)(\hat{\eta  }_N)\equiv \hat{M}_N (\hat {\eta}_N  ) ,
\label{defofnewhamJ}\end{equation}
which are defined as follows  
\begin{eqnarray}&&(\textrm{per}_N P)(\hat{\eta  }_N)\equiv 
\hat{P}_N (\hat {\eta}_N  )
=\left[ \bigl(\hat{P}_N (\hat {\eta}_N  )\bigr)_{ij}\right]_{(N+1)\times (N+1)}
\nonumber \\ &=&\left[
\frac 1 {(i+j-N)!}\left.\frac {\partial ^{i+j-N}P (\hat {u}_N  )}{\partial \varepsilon  ^{i+j-N}}
\right|_{\varepsilon  =0}
 \right]_{(N+1)\times(N+1)}\nonumber
\\ &= & \left[ \begin{array}{cccc}
0& \cdots & 0 & P (\eta  _0)\\ \vdots & \begin{turn}{45}\vdots\end{turn}  & 
\begin{turn}{45}\vdots\end{turn} & 
\frac 1{1!}\left.
\frac {\partial  P (\hat {u}_N  )}{\partial \varepsilon  }
\right|_{\varepsilon  =0}  \\
0 &\begin{turn}{45}\vdots\end{turn} & \begin{turn}{45}\vdots\end{turn}
& \vdots\\ P (\eta  _0)&\frac 1{1!}\left.
\frac {\partial  P (\hat {u}_N  )}{\partial \varepsilon  }
\right|_{\varepsilon  =0}&\cdots&
\frac {1}{N!}\left.\frac {\partial ^N P 
(\hat {u}_N  )}{\partial \varepsilon  ^N }
\right|_{\varepsilon  =0} \end{array}
\right],\ P=J,M;\label{newhamJ}
\end{eqnarray}
and a new hereditary recursion operator defined by 
\bea 
&&(\textrm{per}_N \Phi)(\hat{\eta  }_N)\equiv 
\hat{\Phi }_N (\hat {\eta}_N  )
=\left[ \bigl(\hat{\Phi }_N (\hat {\eta}_N  )\bigr)_{ij}\right]_{(N+1)\times (N+1)}
\nonumber \\ &=&\left[
\frac 1 {(i-j)!}\left.\frac {\partial ^{i-j}  \Phi (\hat {u}_N )}
{\partial \varepsilon  ^{i-j}}\right|_{\varepsilon  =0}
 \right]_{(N+1)\times(N+1)}\nonumber
\\ &= &\left[ \begin{array}{cccc} 
\Phi (\eta _0)&0 & \cdots &  0\\  \frac 1 {1!}
\left. \frac {\partial  \Phi (\hat {u}_N  )}{\partial \varepsilon  }
\right|_{\varepsilon  =0} & \Phi (\eta _0) & \ddots  & \vdots \\
\vdots & \ddots &\ddots& 0 \\ \frac 1{N!}\left.
\frac {\partial  \Phi  (\hat {u}_N  )}{\partial \varepsilon  }
\right|_{\varepsilon  =0}  &\cdots & \frac 1 {1!}
\left. \frac {\partial  \Phi (\hat {u}_N  )}{\partial \varepsilon  }
\right|_{\varepsilon  =0}  & \Phi (\eta _0)
 \end{array}\right ] ; \label{newHOs}
\eea 
where $i,j=0,1,\cdots, N$, and $\hat u_N$ is defined by (\ref{multiplescaleperturbationseries}).
These operators will be used to establish our new bi-Hamiltonian formulation
for the triangular systems by (\ref{triangularSysByPer}).
The structures of these operators originate from those 
proposed for single Hamiltonian formulations in \cite{MaF-CSF1996}.
Only a difference is the scale of perturbations. 
In our previous work \cite{MaF-CSF1996}, single-scale perturbations 
were considered, but in this paper, multi-scale perturbations will need to be considered. Like single-scale perturbations,
multi-scale perturbations also guarantees that two operators 
defined by (\ref{newhamJ}) constitute a Hamiltonian pair and the operator defined by (\ref{newHOs}) is hereditary.
The proofs are direct and very similar to those in 
the case of single-scale perturbations \cite{MaF-CSF1996}, although they are rather laborious and much harder
(see \cite{Ma-MAA1998} for a detailed proof).
Obviously, however, new operators still satisfy 
the following coupled condition:
\begin{equation} 
\hat{\Phi }_N=\hat {M}_N  (\hat {J}_N) ^{-1},\ 
\hat {J}_N \hat {\Psi }_N=\hat{\Phi }_N\hat {J}_N ,\ \hat {\Psi }_N=
(\hat{\Phi }_N)^\dag, \label{eq:ccofJandM}
\end{equation} 
where the superscript $\dag$ means to take the adjoint operation.
The existence of the inverse operator $(\hat {J}_N) ^{-1}$ 
is guaranteed by the existence of $J^{-1}$, following the definition 
of $\hat {J}_N^{-1}$ as in (\ref{defofnewhamJ}) and (\ref{newhamJ}).
The coupled condition (\ref{eq:ccofJandM}) 
ensures \cite{Magri-JMP1978} that 
%symmetries determined by $\hat {\Phi}_N$ constitute an Abelian algebra 
%with the commutator operation of vector fields and  
conserved functionals 
recursively determined by $\hat {\Phi}_N$ 
commute with each other
under two Poisson brackets generated by $\hat {J}_N$ and $\hat {M}_N$. 
  
For the triangular system defined by (\ref{triangularSysByPer}),
new Hamiltonian functionals can be chosen as
\bea 
&& (\textrm{per}_N\tilde{H}_{0}^{\textrm{{\small per}}})(\hat \eta _N)
:=\frac 1 {N!}\frac {\partial ^N}{\partial \varepsilon ^N}
\tilde H_0^{\textrm{{\small per}}}(\hat u_N)\Bigl.
\Bigr |_{\varepsilon=0}
={\displaystyle{\sum_{j=0}^{\textrm{{\small {min}}}(m,N)}
\frac {\alpha _j}{(N-j)!}
\frac {\partial ^{N-j}}{\partial \varepsilon ^{N-j}}\tilde H_{i_j}(\hat {u}_N) \Bigl.\Bigr |
_{\varepsilon =0},}} \label{newH_0per}\\ &&
(\textrm{per}_N \tilde{H}_{1}^{\textrm{{\small per}}})(\hat \eta _N):=
\frac 1 {N!}\frac {\partial ^N}{\partial \varepsilon ^N}
\tilde H_1^{\textrm{{\small per}}}(\hat u_N)
\Bigl. \Bigr |_{\varepsilon=0}
={\displaystyle{\sum_{j=0}^{\textrm{{\small {min}}}
(m,N)}\frac {\alpha _j}{(N-j)!}
\frac {\partial ^{N-j}}{\partial \varepsilon ^{N-j}}\tilde H_{i_j+1}
(\hat {u}_N)\Bigl.\Bigr |_{\varepsilon =0}}}, \quad \ \ 
\label{newH_1per}
\eea
where $\tilde{H}_{0}^{\textrm{{\small per}}}$ and $\tilde{H}_{1}^{\textrm{{\small per}}}$ are defined by (\ref{originalH_01per}), and $\hat u_N$ is defined by (\ref{multiplescaleperturbationseries}). 
They will offer the required Hamiltonian functionals in our bi-Hamiltonian formulation 
of the triangular system (\ref{triangularSysByPer}). 
The above crucial form of the Hamiltonian functionals 
are motivated by a study of the perturbation system of the KdV equation \cite{MaF-PLA1996}.  

Now a direct computation can show that
the triangular system (\ref{triangularSysByPer}) 
has the following bi-Hamiltonian formulation
\begin{equation}\hat {\eta }_{Nt}= {\hat J}_N
\frac {\delta (\textrm{per}_N\tilde {H}_{1}^{\textrm{{\small per}}})}
{\delta \hat \eta _N}=
{\hat M}_N \frac {\delta (\textrm{per}_N\tilde {H}_{0}
^{\textrm{{\small per}}})}
{\delta \hat \eta _N},\ \hat {\eta }_{N}=(\eta _0^T,\eta _1^T,\cdots, \eta _N^T)^T.  
\label{bi-HamiltonianformulationforTSs}
\end{equation}  
Here a Hamiltonian pair of ${\hat J}_N$ and $ {\hat M}_N$
is defined by (\ref{defofnewhamJ}) and (\ref{newhamJ}), and 
two Hamiltonian functionals $\textrm{per}_N\tilde{H}_{0}^{\textrm{{\small per}}}$ and 
$\textrm{per}_N\tilde{H}_{1}^{\textrm{{\small per}}}$ are defined by (\ref{newH_0per}) and (\ref{newH_1per}).
The bi-Hamiltonian formulation 
(\ref{bi-HamiltonianformulationforTSs})
is what we intend to establish for the triangular 
system (\ref{triangularSysByPer}). 
It follows that the triangular 
system (\ref{triangularSysByPer})
is a good example of integrable systems.

%There may exist 
%other kinds of bi-Hamiltonian formulations for (\ref{triangularSysByPer}). 

In fact, let us first introduce 
\begin{equation}\left \{ 
\ba {l}  \textrm{per}_N K_n^{\textrm{{\small per}}}
:=( (K_n^{\textrm{{\small per}}}
(\hat u_N))^T\bigl.\bigr|_{\varepsilon =0},
\frac 1{1!}\frac {\partial }{\partial \varepsilon }(K_n^{\textrm{{\small per}}}
(\hat u_N))^T\bigl.\bigr|_{\varepsilon =0},\cdots,
\frac 1{N!}\frac {\partial ^N}{\partial \varepsilon ^N}(K_n^{\textrm{{\small per}}}
(\hat u_N))^T\bigl.\bigr|_{\varepsilon =0} )^T, \ n\ge 1,
%(\textrm{per}_N(K_n^{\textrm{{\small per}}}))_0^T,
%(\textrm{per}_N(K_n^{\textrm{{\small per}}}))_1^T,\cdots,
%(\textrm{per}_N(K_n^{\textrm{{\small per}}}))_N^T)^T,\ 
%(\textrm{per}_N(K_n^{\textrm{{\small per}}}))_i=
%\frac 1{i!}\frac {\partial ^i}{\partial \varepsilon ^i}K_n^{\textrm{{\small per}}}
%(\hat u_N)\Bigl.\Bigr|_{\varepsilon =0},\ 0\le i\le N,\ n\ge 1,
\vspace{2mm}\\ 
\textrm{per}_N {\tilde H}_n^{\textrm{{\small per}}}
%=(\textrm{per}_N{\Phi } )^n  \textrm{per}_N {\tilde {H}}_{0}^{\textrm{{\small per}}}
:=\frac 1{N!}\frac {\partial ^N}{\partial 
\varepsilon ^N} {\tilde H}_n ^{\textrm{{\small per}}}(\hat u_N)\bigr.\bigl|_{\varepsilon=0}
,\ n\ge 0, 
\ea \right. \label{per_NK_n^{per}per_NtildeH_n^{per}}
\end{equation}
where $K_n^{\textrm{{\small per}}}$ and
${\tilde {H}}_{n }^{\textrm{{\small per}}}$ are defined by
(\ref{K_n^{per}H_n^{per}}), and $\hat u_N$ is defined by (\ref{multiplescaleperturbationseries}).
Then it can directly be verified that 
\begin{equation}\left \{ 
\ba {l} \textrm{per}_N K_n^{\textrm{{\small per}}}
=({\hat \Phi }_N) ^{n-1} ( \textrm{per}_N {K}^{\textrm{\small {per}}}),\ n \ge 1,
\vspace{2mm}\\
\textrm{per}_N {\tilde H}_n^{\textrm{{\small per}}}
%=(\textrm{per}_N{\Phi } )^n  \textrm{per}_N {\tilde {H}}_{0}^{\textrm{{\small per}}}
={\displaystyle {\sum_{j=0}^{\textrm{{\small {min}}}
(m,N)}\frac {\alpha _j}{(N-j)!}
\frac {\partial ^{N-j}}{\partial \varepsilon ^{N-j}}\tilde H_{i_j+n}
(\hat {u}_N)\Bigl.\Bigr |_{\varepsilon =0}}}
 ,\ n \ge 0,
\ea \right.
\nonumber \end{equation} 
and further 
\[ \textrm{per}_N K_n^{\textrm{{\small per}}}= 
{\hat  J }_N \frac {\delta (\textrm{per}_N {\tilde H}_n^{\textrm{{\small per}}})}
{\delta \hat \eta _N} = 
{\hat M }_N \frac {\delta (\textrm{per}_N {\tilde H}_{n-1}^{\textrm{{\small per}}})}
{\delta \hat \eta _N}, \ n\ge 1.
\]
Therefore, it follows from 
the bi-Hamiltonian formulation (\ref{bi-HamiltonianformulationforTSs}) that 
$\textrm{per}_N K_n^{\textrm{{\small per}}},\ n\ge 1,$ and $
\textrm{per}_N( {\tilde H}_n^{\textrm{{\small per}}}),\ n\ge 0,$ defined by 
(\ref{per_NK_n^{per}per_NtildeH_n^{per}}), are 
symmetries and conserved functionals of the triangular system (\ref{triangularSysByPer}),
respectively. This implies that 
the triangular system (\ref{triangularSysByPer})
is integrable if we start from a bi-Hamiltonian system (\ref{originalsolitonsystem}). 

Summing up, the above manipulation shows how to inherit the bi-Hamiltonian formulation
and to compute symmetries and conserved functionals 
for the triangular system (\ref{triangularSysByPer})
while taking perturbations for the starting system (\ref{startingPS}).
In the next section, we perform applications of the above formulation 
to four concrete examples, 
in which new bi-Hamiltonian systems in
both $1+1$ dimensions and $2+1$ dimensions will be formulated. 

\section{Illustrative examples}
\setcounter{equation}{0}

Let us consider the KdV equation
\begin{equation} u_t=K(u)=u_{xxx}+6uu_x.\label{KdVequation}\end{equation} 
It is well known that it has a local bi-Hamiltonian formulation
\cite{Magri-JMP1978,Olver-book1993}
\begin{equation}
u_t=K(u)=J\frac {\delta \tilde H_1}{\delta u}=M\frac {\delta \tilde 
H_0}{\delta u},
\end{equation} 
where the Hamiltonian pair and the Hamiltonian functionals are given by
\begin{equation}
J=\partial_x  ,\ M=\partial_x  ^3 +4u\partial_x  +2u_x,\ 
\tilde H_0=\int \frac 12 u^2\,dx,\ \tilde 
H_1=\int (\frac 12 uu_{xx}+u^3)\,dx. 
\label{JMH_0H_1ofKdV}
 \end{equation} 
Therefore, it has infinitely many symmetries and conserved functionals
\begin{equation}
K_n =\Phi ^n u_x,\ \tilde H_n =\int H_n \,dx,\ 
H_n =\int_0^1 u(\Psi ^n u)(\lambda u)\, d\lambda ,\ n \ge 0, 
\label{K_nH_nofKdV}
\end{equation}
where the hereditary recursion operator $\Phi$ and its adjoint operator 
 $\Psi $ read as 
\begin{equation}\Phi (u)=MJ^{-1}=
\partial_x  ^2 +4u +2u_x\partial_x  ^{-1},\ 
\Psi =\Phi ^\dagger =\partial _x ^2+4u-2\partial _x^{-1}u_x,
 \label{PhiofKdV}
\end{equation}  
where $\partial ^{-1}\partial =\partial \partial ^{-1}=1$.
For example, we can obtain 
\begin{equation}
\left \{ \ba {l}
K_2(u)=u_{5x}+10uu_{xxx}+20u_xu_{xx}+30u^2u_x
, \vspace{2mm}\\
H_2(u)= \frac1 2 uu_{4x}+\frac {10}3u^2u_{xx}+\frac 53 uu_x^2+\frac 52 u^4. 
\ea \right.   \label{K_2tildeH_2}\end{equation}
Note that in (\ref{K_nH_nofKdV}) we added $K_0=u_x$ to the Abelian symmetry algebra
$\{K_n \}_{n=1}^{\infty }$ as defined in 
(\ref{generaleschemeforK_nH_n}).

If we choose the original KdV equation as a starting equation, 
taking single-scale perturbations leads to the standard 
perturbation KdV systems \cite{TamizhmaniL-JPA1983}, which 
were proved to be bi-Hamiltonian \cite{MaF-PLA1996}. In what follows, 
we will formulate other examples of integrable couplings for the KdV 
equation, by choosing proper perturbed equations as starting equations  
and taking bi-scale perturbations in the subsection 4.2.
All examples also show that bi-Hamiltonian systems in both $1+1$ and $2+1$ 
dimensions are many and varied. 

{\bf 4.1 The case of single-scale perturbations:}

We take a special perturbed equation 
\begin{equation} 
u_t=K^{\textrm{{\small per}}}(u)=K_1(u)+  \varepsilon K_1(u)=
J\frac {\delta \tilde H_1^{\textrm{\small per}} }{\delta u}=M
\frac {\delta \tilde H_0^{\textrm{\small per}} }{\delta u}
\label{eq:1stperkdv}
\end{equation} 
with two Hamiltonian functionals 
\begin{equation} 
 \tilde H_0^{\textrm{\small per}}=\tilde H_0+\varepsilon \tilde H_0,\
\tilde H_1^{\textrm{\small per}}=\tilde H_1+\varepsilon \tilde H_1
  \end{equation} 
as a starting equation.
Here $K_1=K, J, M, {\tilde H}_0, {\tilde H}_1$ are given by 
(\ref{JMH_0H_1ofKdV}), (\ref{K_nH_nofKdV}) and {\ref{PhiofKdV}).
The first-order perturbation 
\[\hat u_1=\eta_0+\varepsilon \eta _1\] 
yields the following triangular system
\begin{equation} \left \{\ba {l} \eta  _{0t}= K_1(\eta _0)  =
\eta _{0xxx}+6\eta _0\eta _{0x} , \vspace{2mm}\\ 
\eta  _{1t}= K_1'(\eta _0)[\eta _1]+  K_1(\eta _1)
= \eta_{1xxx}+6(\eta _0\eta _1)_x
+  \eta _{0xxx}+6\eta _0\eta _{0x}.
 \ea \right. \label{eq:1stexofkdv} \end{equation} 
According to our scheme of construction in Section 3,
its Hamiltonian pair and  corresponding hereditary recursion operator are 
\bea &&
\hat {J }_1 =\left [\ba {cc} 0
& \partial_x  \vspace{2mm} \\ \partial_x  & 0
 \ea \right],\ \hat {M }_1 = \left [\ba {cc} 0
& {M}_{0}\vspace{2mm}\\
 {M}_{0}& {M}_{1}
\ea \right ], \ \hat {\Phi }_1=\left [\ba {cc} 
\Phi_{0}& 0\vspace{2mm}\\
\Phi_{1}& \Phi_{0}
\ea \right], \eea 
with the entries of $\hat M_1$ and $\hat \Phi _1$ being defined by 
\begin{equation} 
 {M}_{i}=\delta _{i0}\partial_x  ^3 +4\eta _0 \partial_x  +2 \eta _{0x} 
,\   {\Phi }_{i}=\delta _{i0}\partial_x  ^2 
+4\eta _i +2\eta _{ix}\partial_x  ^{-1},
\ i=0,1,\end{equation} 
where $\delta _{0i}$ is the Kronecker symbol. 
The triangular system (\ref{eq:1stexofkdv}) has a local 
bi-Hamiltonian formulation:
\begin{equation} \hat {\eta }_{1t}=
\hat {J}_1
\frac {\delta (\textrm{per} _1\tilde {H}^{\textrm{{\small per}}}_{1} )}
{\delta \hat {\eta  }_1}= \hat{M}_1
\frac {\delta (\textrm{per}_1\tilde {H}^{\textrm{{\small per}}}_{0})}
{\delta \hat {\eta   }_1},\ \hat {\eta   }_1=(\eta _0,\eta _1)^T,
\end{equation} 
with two Hamiltonian functionals 
\begin{equation}\left\{
\ba {l} \displaystyle \textrm{per}_1 \tilde H_{0}
^{\textrm{{\small {per} }} }
=(\textrm{per}_1 \tilde H_{0}
^{\textrm{{\small {per}}}})(\hat {\eta }_1)=
 \frac {\partial  \tilde H_0(\hat {u} _1)}{\partial \varepsilon }
\Bigr.\Bigl|_{\varepsilon =0}+ \tilde  H_0(\eta _0)
 =\int (\eta_0\eta _1+\frac 12   \eta _0^2)\,dx ,
\vspace{2mm} \\ 
{\displaystyle \textrm{per}_1 \tilde H_{1}^{\textrm{{\small {per}}}}
=(\textrm{per}_1 \tilde H_{1}^{\textrm{{\small {per}}}})(\hat {\eta }_1) 
=\frac {\partial \tilde H_1(\hat {u} _1)}{\partial \varepsilon }
\Bigr.\Bigl|_{\varepsilon =0}+ \tilde H_1(\eta_0 )}
 \vspace{2mm}\\
\qquad \qquad ={\displaystyle
 \int (\frac1 2 \eta_{0xx}\eta _1+ \frac1 2 \eta_{0}\eta _{1xx}+
3\eta _0^2\eta _1+\frac 12 \eta _0\eta _{0xx}+\eta _0^3)\,dx }.
\ea \right.\end{equation} 
Noting that in this example, we have 
\[ { K}_n  ^{\textrm{{\small {per}}}}={ K}_n +\varepsilon { K}_n ,\ 
 {\tilde H}_n  ^{\textrm{{\small {per}}}}={\tilde H}_n +\varepsilon {\tilde H}_n,\ n\ge 0, \]
infinitely many symmetries and conserved functionals
of the triangular system (\ref{eq:1stexofkdv}) are computed as follows
\begin{equation}\left \{ \ba {l} 
\textrm{per}_1( K_n^{\textrm{\small per}})
%=(\textrm{per}_1 {\Phi }) ^{n }\textrn{per}_1 {K}^{\textrm{{\small per}}}
=\left[\ba {c} K_{n }(\eta _0)\vspace{2mm} \\
\frac {\partial K_{n }(\hat{u}_1)}{\partial \varepsilon }
\Bigr.\Bigl|_{\varepsilon =0} +  K_{n }(\eta _0) \ea \right] ,\
n \ge 0,
\vspace{2mm} \\ 
\textrm{per}_1({\tilde H}_n ^{\textrm{\small per}})
%=\hat H_{n 1}^{\textrm{{\small per}}}
=\frac {\partial {\tilde H}_n (\hat{u}_1)}{\partial \varepsilon }
\Bigr.\Bigl|_{\varepsilon =0} +  {\tilde H}_n (\eta _0 ),
\ n \ge 0. \ea \right.\end{equation} 

Secondly, we take another special perturbed equation
\begin{equation}u_t=K^{\textrm{{\small per}}}(u)
=K_1(u)+  \varepsilon K_2(u),  \end{equation} 
as a starting equation, which can be written as 
a bi-Hamiltonian system 
\begin{equation}
u_t=K^{\textrm{{\small per}}}(u)=J\frac {\delta \tilde H_1
^{\textrm{{\small per}}} }{\delta u}
=M\frac {\delta \tilde H_0^{\textrm{{\small per}}} }
{\delta u},\ \tilde H_0^{\textrm{{\small per}}}=\tilde H_0+\varepsilon 
\tilde H_1,\ 
 \tilde H_1^{\textrm{{\small per}}}=\tilde H_1+\varepsilon \tilde H_2.
\end{equation}
Here $K_1=K$,
$J, M, {\tilde H}_0, {\tilde H}_1,{\tilde H}_2, K_2, $ 
are determined by 
(\ref{KdVequation}), (\ref{JMH_0H_1ofKdV}) and (\ref{K_2tildeH_2}).
The second-order perturbation yields the following triangular system:
\begin{equation} \left \{\ba {l} \eta  _{0t}= \eta  _{0xxx}
+6\eta  _0\eta  _{0x},\vspace{2mm}\\ 
\eta  _{1t}=  \eta _{1xxx}+6(\eta _0\eta _1)_x+
  K_2(\eta _{0}),
\vspace{2mm}\\ \eta  _{2t}=\eta _{2xxx}
+6(\eta _0\eta _2)_x+6\eta _1\eta _{1x}+  
\frac {\partial K_2(\hat u_2)}
{\partial \varepsilon}\Bigl.\Bigr|_{\varepsilon =0}, \ea \right. 
\label{eq:2ndexofkdv}
\end{equation} 
where 
\begin{eqnarray}&&
\frac {\partial K_2(\hat u_2)}
{\partial \varepsilon}\Bigl.\Bigr|_{\varepsilon =0}
=\eta _{1,5x}
+10\eta _1\eta _{0xxx}+10\eta _0\eta _{1xxx}
\vspace{2mm}\nonumber \\
&& \quad \qquad +20\eta _{1x}\eta _{0xx}+20\eta _{0x}\eta _{1xx}
+60\eta _0\eta _{0x}\eta _1
+30\eta _0^2\eta _{1x}. 
 \end{eqnarray} 
The corresponding Hamiltonian pair and hereditary recursion operator are 
\begin{equation}
\hat {J }_2 =\left [\ba {ccc} 0&0
& \partial_x  \vspace{2mm} \\ 0& \partial_x  & 0 \vspace{2mm}\\
\partial_x  & 0 &0
 \ea \right],\ \hat {M }_2 = \left [\ba {ccc} 0&0
& {M}_{0}\vspace{2mm}\\ 0& {M}_{0}& {M}_{1}\vspace{2mm}\\
 {M}_{0}& {M}_{1}& {M}_{2}\ea \right ], \ 
\hat {\Phi }_2=\left [\ba {ccc} \Phi_{0}& 0&0\vspace{2mm}\\
\Phi_{1}& \Phi_{0}&0\vspace{2mm}\\
\Phi_{2}& \Phi_{1}&\Phi_{0}\ea \right],\end{equation} 
where the entries of $\hat M_2$ and $\hat \Phi _2$ are given by
\begin{equation} 
 {M}_{i}=\delta _{i0}\partial_x  ^3 +4\eta _0 \partial_x  +2 \eta _{0x} 
,\   {\Phi }_{i}=\delta _{i0}\partial_x  ^2 
+4\eta _i +2\eta _{ix}\partial_x  ^{-1},
\ 0\le i\le 2. \end{equation} 
Through our scheme of construction in Section 3, the triangular system (\ref{eq:2ndexofkdv}) has a local bi-Hamiltonian formulation
\begin{equation} \hat {\eta }_{2t}=
\hat {J}_2
\frac {\delta (\textrm{per}_2\tilde {H}^{\textrm{{\small per}}}_{1})}
{\delta \hat {\eta  }_2}=
\hat{M}_2
\frac {\delta (\textrm{per}_2\tilde {H}^{\textrm{{\small per}}}_{0})}
{\delta \hat {\eta  }_2},\ \hat {\eta }_2=(\eta _0,\eta _1,\eta _2)^T,
\end{equation} 
with two Hamiltonian functionals 
\begin{equation}\left\{
\ba {rcl}{\displaystyle \textrm{per}_2 \tilde 
H_{0}^{\textrm{\small {per}}}}
&=&(\textrm{per}_2 \tilde H_{0}^{\textrm{\small {per}}})(\hat {\eta }_2)=
\frac1 2 \frac {\partial ^2 \tilde 
H_0(\hat {u}_2)}{\partial \varepsilon ^2}
\Bigr.\Bigl|_{\varepsilon =0} +  \frac {\partial \tilde 
H_1(\hat {u}_2)}{\partial 
\varepsilon }\Bigr.\Bigl|_{\varepsilon =0} \vspace{2mm}\\
&=&{\displaystyle \int 
 (\frac 12 \eta _1^2+\eta _0\eta _2+  \frac 12 \eta _{0xx}\eta _1 +
\frac 12 \eta _0\eta _{1xx}+3\eta _0^2\eta _1)}\,dx ,
\vspace{2mm} \\
{\displaystyle \textrm{per}_2 \tilde H_{1}^{\textrm{\small {per}}}}
&=&(\textrm{per}_2 \tilde H_{1}^{\textrm{\small {per}}})(\hat {\eta }_2)
 =\frac1 2 \frac {\partial ^2 \tilde 
H_1(\hat {u}_2)}{\partial \varepsilon ^2}
\Bigr.\Bigl|_{\varepsilon =0} +  \frac {\partial \tilde 
H_2(\hat {u}_2)}{\partial 
\varepsilon }\Bigr.\Bigl|_{\varepsilon =0} \vspace{2mm}\\
&=&{\displaystyle \int (
\frac 12 \eta _{0xx}\eta _2+\frac 12 \eta _1\eta _{1xx}+\frac 12
\eta _0\eta _{2xx}}
 +3\eta _0\eta _1^2+ 3\eta _0^2\eta _2 )\,dx + 
\frac {\partial \tilde H_2(\hat {u}_2)}{\partial \varepsilon }
\Bigr.\Bigl|_{\varepsilon =0} ,
\ea \right.\end{equation} 
where \begin{eqnarray}&&  
\frac {\partial \tilde H_2(\hat {u}_2)}{\partial \varepsilon }
\Bigr.\Bigl|_{\varepsilon =0} ={\displaystyle
\int ( \frac 12 \eta _{0,4x}\eta _1+\frac 12 \eta _0\eta _{1,4x}
+\frac 53\eta _{0x}^2\eta _1 }\nonumber \vspace{2mm}\\ && 
\quad  \qquad\quad  +\frac {10}3 \eta _0\eta _{0x}\eta _{1x}+\frac {20}3
 \eta _0\eta _{0xx}\eta _{1}+\frac {10}3
  \eta _0^2\eta _{1x}+10\eta_0^3\eta _1 )\,dx .  \end{eqnarray} 
Its infinitely many symmetries and  conserved functionals read as
\begin{equation} \left \{\ba {l}
\textrm{per}_2( K_n^{\textrm{\small per}})=
%\hat {\Phi }_2 ^n   \hat {K}^{\textrm{{\small per}}}_2 =
\left [\ba {c} K_{n }(\eta _0)\vspace{2mm}\\ \frac {\partial K_{n }
(\hat {u}_2)}{\partial \varepsilon}\Bigl.\Bigr|_{\varepsilon =0}+K_{n +1} (\eta _0), \vspace{2mm} \\
\frac 12 \frac {\partial ^2K_{n } (\hat {u}_2)}{\partial \varepsilon ^2}
\Bigl.\Bigr|_{\varepsilon =0} +\frac {\partial K_{n +1}
(\hat {u}_2)}{\partial \varepsilon}
\Bigl.\Bigr|_{\varepsilon =0}  \ea \right ],\ n \ge 0,\vspace{2mm}\\
\textrm{per}_2  {\tilde H }_ {n }^{\textrm{{\small per}}}=
%\hat {H}_{n 2}^{\textrm{{\small per}}}=
\frac 12 \frac {\partial ^2{\tilde H}_{n }
(\hat {u}_2)}{\partial \varepsilon ^2}
\Bigl.\Bigr|_{\varepsilon =0}+
\frac {\partial {\tilde H}_{n +1}
(\hat {u}_2)}{\partial \varepsilon}
\Bigl.\Bigr|_{\varepsilon =0},\ 
n \ge 0,\ea \right.
\end{equation} 
by using 
${K}_{n }^{\textrm{{\small per}}}={K}_n +\varepsilon 
{K}_{n +1}$ and
${\tilde H}_{n }^{\textrm{{\small per}}}={\tilde H}_n +\varepsilon 
{\tilde H}_{n +1}$ 
in this example.

{\bf 4.2 The case of bi-scale perturbations:}

We would like to exhibit two examples in the case of bi-scale perturbations
and show that  
multi-scale perturbations can lead to bi-Hamiltonian systems
in higher spatial dimensions. A concrete example in $2+1$
dimensions for the KdV equation is the following triangular system
\begin{equation}\left \{\begin{array}{l}
\eta  _{0t_1}=\eta  _{0xxx}+6\eta  _0\eta  _{0x},\vspace{2mm} \\
\eta  _{1t_1}=\eta  _{1xxx}+3\eta  _{0xxy}+6(\eta  _0\eta  _{1})_x+6\eta  _0\eta  _{0y},
\end{array}\right.\label{eq:3rdexofkdv}\end{equation} 
resulting from 
%a starting equation of 
the KdV equation (\ref{KdVequation})
by a first-order bi-scale perturbation 
\begin{equation}
 \hat {u}_1=\displaystyle{\eta _0(t,x,y)+\varepsilon 
\eta _1(t,x,y)},\ y=\varepsilon x.\end{equation} 
This systems was furnished in \cite{MaF-PLA1996}, and
based on our scheme of construction in Section 3, 
it has the following local bi-Hamiltonian formulation
\begin{equation} \hat {\eta }_{1t}=
\hat {J}_1
\frac {\delta (\textrm{per}_1\tilde {H}_{1})}{\delta \hat {\eta  }_1}=
\hat{M}_1 \frac {\delta (\textrm{per}_1\tilde {H}_{0})}
{\delta \hat {\eta  }_1}, \ \hat {\eta }_1=(\eta _0,\eta _1)^T, 
\label{first2+1localbiHamiltoniansystem} \end{equation} 
where the Hamiltonian pair reads as 
\begin{equation} \hat J_1=\left[\ba {cc}0& \partial _x \vspace{2mm}\\
\partial _x &\partial _y\ea \right],\ 
\hat M_1=\left[\ba {cc} 0 & \partial _x ^3+2\eta _{0x}+4\eta _0\partial _x
\vspace{2mm}\\
\partial _x ^3+2\eta _{0x}+4\eta _0\partial _x & Q \ea \right],
\end{equation}
with $Q$ being defined by 
\begin{equation}
Q =\frac {\partial }{\partial \varepsilon }M(\hat u_1)
\Bigl.\Bigl|_{\varepsilon=0}
= 3\partial _x^2\partial _y +2\eta _{1x}+2\eta_{0y}+4\eta _1\partial _x
+4\eta _0\partial _y,\end{equation}
 and the Hamiltonian functionals are 
\begin{equation} \left\{\ba {l} \textrm{per}_1 \tilde H_{0} =
\frac {\partial }{\partial \varepsilon }\tilde H_0(\hat u_1)
\Bigl.\Bigl|_{\varepsilon=0} 
=\iint \eta _0\eta _1 \, dxdy, \vspace{2mm}\\
 \textrm{per}_1\tilde H_{1}=
\frac {\partial }{\partial \varepsilon }\tilde H_1(\hat u_1)
\Bigl.\Bigl|_{\varepsilon=0}=
\iint (\frac 12 \eta _0\eta _{1xx}+\eta _0\eta _{0xy}
+\frac 12 \eta _1\eta _{0xx}+3\eta _0^3\eta _1)\, dxdy .
\ea \right. \end{equation} 
Moreover, the above Hamiltonian pair yields
a hereditary recursion operator in $2+1$ dimensions
\begin{equation}\hat {\Phi }_{1} = 
\left[\begin{array} {cc}
 \partial _x^2 +2\eta _{0x}\partial _x^{-1} +4\eta _0& 0
\vspace{2mm} \\
2\partial _x\partial _y-2\eta _{0x}\partial _x^{-2}\partial _y
+2(\eta _{1x}+\eta _{0y})\partial _x^{-1}+4\eta _1
&  \partial _x^2 +2\eta _{0x}\partial _x^{-1} +4\eta _0
 \end{array} \right],
\end{equation} 
for the triangular system (\ref{eq:3rdexofkdv}).

The second example is the following
\begin{equation}
\left \{\begin{array}{rcl}
\eta  _{0t_1}&=&K_1(\eta _0)=\eta  _{0xxx}+6\eta  _0\eta  _{0x},\vspace{2mm} \\
\eta  _{1t_1}&=&\frac{\partial K_1(\hat {u}_2)}
{\partial \varepsilon }\Bigr.\Bigl|_{\varepsilon =0}
 +K_1(\eta _0)\vspace{2mm}  \\
& =&\eta  _{1xxx}+3\eta  _{0xxy}+6(\eta  _0\eta  _{1})_x+6\eta  _0\eta  _{0y}+
\eta  _{0xxx}+6\eta  _0\eta  _{0x},\vspace{2mm}  \\
\eta  _{2t_1}&=&
\frac 1{2!}\frac{\partial ^2K_1(\hat {u}_2)}{\partial \varepsilon ^2}
\Bigr.\Bigl|_{\varepsilon =0}
+\frac{\partial K_1(\hat {u}_2)}{\partial \varepsilon }
\Bigr.\Bigl|_{\varepsilon =0}\vspace{2mm}  \\
& =&\eta _{2xxx}+3\eta _{1xxy}+3\eta _{0xyy}+6(\eta _0\eta _2)_x
+6\eta _1\eta _{1x}+6(\eta _0\eta _1)_y \vspace{2mm}  \\ &&
+\eta  _{1xxx}+3\eta  _{0xxy}+6(\eta  _0\eta  _{1})_x+6\eta  _0\eta  _{0y},
\end{array}\right.\label{eq:4thexofkdv}
\end{equation}
which can be generated from a perturbed KdV equation (\ref{eq:1stperkdv})
under the second-order bi-scale perturbation 
\begin{equation} \hat {u}_2=\displaystyle{\eta (t,x,y)+ \varepsilon 
\eta _1(t,x,y)+ \varepsilon \eta _2(t,x,y)},\ y=\varepsilon x .\end{equation}
According to our scheme of construction in Section 3, 
the corresponding Hamiltonian pair and recursion operator read as 
 \begin{equation} \hat {J}_2  = \left[\ba {ccc}0& 0&\partial _x \vspace{2mm}\\
0&\partial _x &\partial _y \vspace{2mm}\\
\partial _x &\partial _y & 0 \ea \right],\ 
\hat M_2 =\left[\ba {ccc} 0 & 0&M_0 \vspace{2mm}\\
0&M_0 & M_1 \vspace{2mm}\\ M_0&M_1&M_2
\ea \right],\ 
 \hat {\Phi }_{2}= 
\left[\begin{array} {ccc} 
\Phi _0& 0 &0
\vspace{2mm} \\
\Phi_1 & \Phi _0& 0\vspace{2mm} \\
\Phi_2 & \Phi _1& \Phi _0
 \end{array} \right],\label{eq:defofhatJ_2hatM_2hatPhi_2}
\end{equation}
where the entries of $\hat {M}_2$ are defined by
\begin{equation}\left \{ \ba {l}
M_0=M (\hat {u}_2)|_{\varepsilon =0}
=\partial _x ^3+2\eta _{0x}+4\eta _0\partial _x
 \vspace{2mm}\\
M_1 = \frac 1{1!}\frac {\partial M (\hat {u}_2)}{\partial \varepsilon }
\Bigl.\Bigr|_{\varepsilon =0}
=3\partial _x^2\partial _y +2\eta _{1x}+2\eta_{0y}+4\eta _1\partial _x
+4\eta _0\partial _y,\vspace{2mm}\\
M_2=
\frac 1{2!}\frac {\partial ^2 M (\hat {u}_2)}{\partial \varepsilon ^2}
\Bigl.\Bigr|_{\varepsilon =0}
=3\partial _x\partial _y^2+
2\eta _{2x}+2\eta _{1y}+ 4\eta _2\partial _x+4\eta _1\partial _y,
\ea \right. \end{equation}
and the entries of $\hat {\Phi }_2$, by
\begin{equation}
\left\{\ba {l}
\Phi _0=\Phi (\hat {u}_2)|_{\varepsilon =0}
= \partial _x^2 +2\eta _{0x}\partial _x^{-1} +4\eta _0,\vspace{2mm}\\
\Phi _1=\frac 1{1!}\frac {\partial \Phi (\hat {u}_2)}{\partial \varepsilon }
\Bigl.\Bigr|_{\varepsilon =0}
=2\partial _x\partial _y
+2(\eta _{1x}+\eta _{0y})\partial _x^{-1}
-2\eta _{0x}\partial _x^{-2}\partial _y
+4\eta _1,\vspace{2mm}\\
\Phi _2=\frac 1{2!}\frac {\partial ^2\Phi (\hat {u}_2)}
{\partial \varepsilon ^2}
\Bigl.\Bigr|_{\varepsilon =0}
=\partial _y^2+2(\eta _{2x}+\eta _{1y})\partial _x^{-1}
-2(\eta _{1x}+\eta _{0y})
\partial _{x}^{-2}\partial _y
+2\eta _{0x}\partial _x^{-3}\partial _y^2+4\eta _2.
\ea \right . \end{equation}
The 2+1 triangular system (\ref{eq:4thexofkdv})
has a local bi-Hamiltonian formulation 
\begin{equation} \hat {\eta }_{2t}=
\hat {J}_2
\frac {\delta (\textrm{per}_2 \tilde {H}_{1})}{\delta \hat {\eta  }_2}=
\hat{M}_2
\frac {\delta (\textrm{per}_2 \tilde {H}_{0})}
{\delta \hat {\eta  }_2},
\ \hat {\eta }_2=(\eta _0,\eta _1,\eta _2)^T,
\label{second2+1localbiHamiltoniansystem}\end{equation}
with 
a Hamiltonian pair $\hat {J}_2$ and $\hat {M}_2$ 
being defined by (\ref{eq:defofhatJ_2hatM_2hatPhi_2}), and 
two Hamiltonian functionals, by 
\begin{equation}\left\{\ba {l}
\textrm{per}_2 \tilde {H}^{\textrm{\small per}}_{0}=
\frac 1 {2!}\frac {\partial ^2\tilde {H}^{\textrm{\small per}}_0}
{\partial \varepsilon ^2}\Bigr.\Bigl|_{\varepsilon =0} =
\frac 1 {2!}\frac {\partial ^2\tilde {H}_0}{\partial \varepsilon ^2}
\Bigr.\Bigl|_{\varepsilon =0}
+\frac {\partial \tilde {H}_0}{\partial \varepsilon }
\Bigr.\Bigl|_{\varepsilon =0}
=\iint (\frac 12 \eta _0\eta _2+\frac 14 \eta _1^2+\eta _0\eta _1)dxdy ,
\vspace{2mm}\\
\textrm{per}_2 \tilde {H}^{\textrm{\small per}}_{1}=
\frac 1 {2!}\frac {\partial ^2\tilde {H}^{\textrm{\small per}}_1}
{\partial \varepsilon ^2}\Bigr.\Bigl|_{\varepsilon =0}=
\frac 1 {2!}\frac {\partial ^2\tilde {H}_1}{\partial \varepsilon ^2}
\Bigr.\Bigl|_{\varepsilon =0}
+\frac {\partial \tilde {H}_1}{\partial \varepsilon }\Bigr.\Bigl|_{\varepsilon =0}
\vspace{2mm}\\ \qquad \qquad \quad 
=\iint \bigl[\frac 14( \eta _0\eta _{2xx}+\eta _1\eta _{1xx}+\eta _2\eta _{0xx}
+2\eta _0\eta _{1xx}+2\eta _1\eta _{0xx})\vspace{2mm}\\
\qquad \qquad \quad 
+\frac 12 (\eta _0\eta _{1xy}+\eta _1\eta _{0xy}+2\eta _0\eta _{0xy})
+\frac 14 \eta _0\eta _{0yy}+3(\eta _0\eta _1^2+\eta _0^2\eta _1
+\eta _0^2\eta _2)\bigr]dxdy .\ea \right. 
\end{equation} 

Both $2+1$ dimensional triangular systems above have infinitely many 
symmetries and conserved functionals due to the existence of hereditary 
recursion operators, and thus they are also integrable in the sense of 
the existence of the Abelian symmetry algebra \cite{Fokas-SAM1987}.  
Note that under the bi-scale perturbation
\[ \hat {u}_N= \sum_{i=0}^N \varepsilon ^i\eta _i(t,x,y),\ y=\varepsilon x,
\] 
we have, for example, 
\[ \partial _x\to \partial _x+ \varepsilon \partial _y,\ 
\hat {u}_{Nx}\to \sum _{i=0}^N\varepsilon ^i (\eta _{ix}+\varepsilon 
\eta _{iy}),\ 
\hat {u}_{Nxx}\to \sum_{i=0}^N\varepsilon ^i(\eta _{ixx}+2\varepsilon 
\eta _{ixy}+\varepsilon ^2\eta _{iyy} ).
\]
These equalities have been used in the above deduction of bi-Hamiltonian systems
in $2+1$ dimensions.

\section{Concluding remarks}
\setcounter{equation}{0}

We have proposed a bi-Hamiltonian formulation (\ref{bi-HamiltonianformulationforTSs}) 
for the triangular systems (\ref{triangularSysByPer}) resulted by perturbations around
solutions of the perturbed systems. The symmetry problem can lead to 
a special case (\ref{symmetrysystem}) of our triangular systems 
(\ref{triangularSysByPer}), which is generated by the 
first-order perturbation. 
However, the perturbation system (\ref{symmetrysystem}) is a little more 
general than the symmetry problem itself. It is because 
the second component system of the perturbation system
(\ref{symmetrysystem}) needs to hold only for 
a solution of the original system $u_t=K(u)$, 
but the same system in the symmetry problem needs to hold
for all solutions of $u_t=K(u)$.
The resulting formulation gives a way to construct 
various integrable couplings in both lower dimensions and higher dimensions 
for bi-Hamiltonian systems, 
all of which at least possess infinitely many commuting symmetries and conserved functionals. 
Four illustrative examples were given for the KdV equation,
which contain two $2+1$ dimensional local bi-Hamiltonian systems
(\ref{first2+1localbiHamiltoniansystem}) and (\ref{second2+1localbiHamiltoniansystem}).
%To produce examples, all what we need to do is to perform direct computations.
%We remark that there may exist other kinds of 
%bi-Hamiltonian formulations for the triangular systems (\ref{triangularSysByPer}).

The triangular system (\ref{eq:3rdexofkdv}) was first introduced in 
\cite{MaF-PLA1996}, whose Painlev\'e property and zero curvature representation were discussed by Sakovich \cite{Sakovich-JNMP1998}.
General triangular systems resulted by multi-scale perturbations 
also can possess rich structures of zero curvature representations. 
If multi-scale perturbations are taken into account, the involved spectral
parameters, denoted by $\mu _i,\ 0\le i\le N,$ may vary with respect to 
the spatial variables \cite{Sakovich-JNMP1998,Ma-MAA1998}, although they
need to satisfy some conditions, for example,
 \[ \mu_{0x}=0,\ \mu _{ix}+\mu _{i-1,y}=0,\ 1\le i\le N,\]   
in the case of bi-scale perturbations
\[ \hat {u}_N=\sum_{i=0}^N \varepsilon ^i\eta_i (x,y,t)=
 \sum_{i=0}^N \varepsilon ^i\eta _i(x,\varepsilon x,t).
 \]
More interestingly, our $2+1$ dimensional bi-Hamiltonian systems 
(\ref{first2+1localbiHamiltoniansystem}) and (\ref{second2+1localbiHamiltoniansystem})
are local and possess hereditary recursion operators,
and thus they enjoy
a different feature from known scalar integrable equations in $2+1$
dimensions.
To our best knowledge, (\ref{first2+1localbiHamiltoniansystem}) and (\ref{second2+1localbiHamiltoniansystem})
 are the first two examples of local $2+1$ dimensional bi-Hamiltonian systems 
with hereditary recursion operators.
They also can provide useful information for classifying 
integrable systems in $2+1$ dimensions by the symmetry approach
\cite{MikhailovSS-book1991}.

We remark that our general bi-Hamiltonian formulation in Section 3 can be used to 
establish bi-Hamiltonian formulations for a hierarchy of coupled KdV systems 
introduced in \cite{Ma-JPA1998}, although it does not work 
for the other two hierarchies of coupled KdV systems
furnished in \cite{Ma-masterthesis1985,AntonowiczF-PD1987} and \cite{MaP-PLA1998}. Moreover, 
our triangular systems, especially (\ref{specificIC}),
starting from the KdV equation, provide examples of bi-Hamiltonian 
systems among the integrable coupled KdV systems described by G\"urses and Karasu
\cite{GuersesK-JMP1998}, and general triangular systems can provide new 
bi-Hamiltonian systems of other types, for example, bi-Hamiltonian systems 
of coupled fifth KdV equations and coupled modified KdV equations.
Nonlinearization resulting from symmetry constraints also 
can be manipulated for linking our triangular systems to 
finite-dimensional integrable Hamiltonian systems \cite{MaZ-PA2001}. 

We finally point out that our general scheme requires
a bi-Hamiltonian structure of the starting system.
%What we proved is to show how to 
Nevertheless, we can still make the perturbation to get triangular
systems from non-Hamiltonian systems such as the KP hierarchy,
and study integrable properties for the resulting triangular systems.

\vskip 1mm
\noindent{\bf Acknowledgments:} 
The work was supported by 
the City University of Hong Kong (SRGs 7001041, 7001178)
and 
the Research Grants Council of Hong Kong (CERG 9040466).
The author is also grateful to the referee for helpful comments.

\setlength{\baselineskip}{15pt}


\small
\begin{thebibliography}{99}

\bibitem{Magri-JMP1978} F. Magri,
 {J. Math. Phys.}
{\bf 19}, 1156--1162 (1978).
\bibitem{CaseR-JMP1981} K. M. Case and A. M. Roos,
 {J. Phys. A: Math. Gen.} {\bf 22}, 2824--2830 (1981).
\bibitem{LakshmananT-JMP1985} M. Lakshmanan and K. M. Tamizhmani, 
{J. Math. Phys.} {\bf 26}, 1189--1200 (1985).
\bibitem{MaF-CSF1996} W. X. Ma and B. Fuchssteiner, 
{Chaos, Solitons $\&$ Fractals} {\bf 7}, 1227--1250 (1996).
\bibitem{Fuchssteiner-book1993} B. Fuchssteiner,
  in: {\it Applications of Analytic and Geometric Methods to Nonlinear 
  Differential Equations}  
  NATO Advanced Science Institutes Series C: 
  Mathematical and Physical Sciences 413, 
 ed Clarkson P A (Dordrecht: Kluwer) pp 125--138 (1993).
\bibitem{Ma-MAA1998} W. X. Ma, 
%{\it Integrable couplings of soliton 
% equations by perturbations I. A general theory and application to the KdV hierarchy}, 
{Methods Appl. Anal.}, {\bf 7}, 21--56 (2000).
\bibitem{GelfandD-FAA1979} I. M. Gelfand and I. Ya Dorfman,
 {Func. Anal. Appl.} {\bf 13}, 248--262 (1979).
\bibitem{FokasF-LNC1980} A. S. Fokas and B. Fuchssteiner,
 {Lett. Nuovo Cimento} {\bf 28}, 299--303 (1980).
\bibitem{Blaszak-book1998} M. Blaszak,
{\it Multi-Hamiltonian Theory of Dynamical Systems}
(Berlin: Springer-Verlag, 1998).
\bibitem{Olver-book1993} P. J. Olver, {\it Applications of Lie Groups
  to Differential Equations} Second edition, 
  Graduate Texts in Mathematics 107 
(New York: Springer-Verlag, 1993).
\bibitem{FuchssteinerF-PD1981} B. Fuchssteiner and A. S. Fokas,
  {Physica D} {\bf 4}, 47--66 (1981).
\bibitem{Fuchssteiner-NATMA1979} B. Fuchssteiner,
{ Nonlinear Analysis TMA} {\bf 3}, 849--862 (1979).
\bibitem{Ma-JPA1990} W. X. Ma, {J. Phys. A: Math. Gen.} {\bf 23}, 2707--2716 (1990).
\bibitem{Ma-SCA1991} W. X. Ma, 
{Science in China A} {\bf 34}, 769--782 (1991).
\bibitem{TamizhmaniL-JPA1983}K. M. Tamizhmani and M. Lakshmanan,
{ J. Phys. A: Math. Gen.} {\bf 16}, 3773--3782 (1983).
\bibitem{MaF-PLA1996} W. X. Ma and B. Fuchssteiner,
  {Phys. Lett. A} {\bf 213}, 49--55 (1996).
\bibitem{Fokas-SAM1987} A. S. Fokas,
{ Studies in Appl. Math.} {\bf 77}, 253--299 (1987).
\bibitem{Sakovich-JNMP1998}S. Yu Sakovich,
{ J. Nonlinear Math. Phys.} {\bf 5}, 230--233 (1998).
\bibitem{MikhailovSS-book1991} A. V. Mikhailov, A. B. Shabat and  V. V.
  Sokolov, in: 
  {\it What is integrability?} Springer Series in Nonlinear 
  Dynamics, ed Zakharov V E (Berlin: Springer) 
pp 115--184 (1991).
\bibitem{Ma-JPA1998} W. X. Ma, {J. Phys. A: Math. Gen.} 
{\bf 31}, 7585--7591 (1998).
\bibitem{Ma-masterthesis1985} W. X. Ma,
{\it On the Generalized Hamiltonian Structure of Nonlinear Evolution Equations},
Master Thesis, Academia Sinica, Beijing, 1985.
\bibitem{AntonowiczF-PD1987} M. Antonowicz and A. P. 
Fordy, 
{ Physica D} {\bf 28}, 345--357 (1987).
\bibitem{MaP-PLA1998} W. X. Ma and M. Pavlov, {Phys. Lett. A} 
  {\bf 28}, 511--522 (1998).
\bibitem{GuersesK-JMP1998} M. G\"urses and A. Karasu,
{J. Math. Phys.}
  {\bf 39}, 2103--2111 (1998).
\bibitem{MaZ-PA2001} W. X. Ma and R. G. Zhou,
%  {\it Nonlinearization of spectral problem for the perturbation KdV system},
 {Physica A} {\bf 296}, 60--74 (2001). 

\end{thebibliography}
\end{document}